\documentclass[runningheads,a4paper,envcountsame]{elsarticle}
\usepackage[utf8]{inputenc} 
\usepackage[T1]{fontenc} 
\usepackage{lmodern}
\usepackage[english]{babel} 
\usepackage{amsmath,amssymb,amsfonts,amsthm}

\usepackage{array}
\usepackage{enumerate} 
\usepackage{fancyhdr}
\usepackage{arcs}
\usepackage{tabvar}
\usepackage{mathrsfs}
\usepackage{dsfont}
\usepackage{stmaryrd}
\usepackage{geometry}
\usepackage{tikz}
\usetikzlibrary{shapes,arrows,calc,decorations.pathmorphing,decorations.pathreplacing}
\renewcommand*{\hat}[1]{\widehat{#1}}

\newcounter{config}

\newcounter{regle}

\newtheorem{thm}{Theorem}
\newtheorem{conj}[thm]{Conjecture}
\newtheorem{prop}[thm]{Proposition}
\newtheorem{remark}[thm]{Remark}
\newdefinition{problem}[thm]{Problem}
\newdefinition{question}[thm]{Question}
\newdefinition{definition}[thm]{Definition}

\begin{document}
\begin{frontmatter}
  \title{A Brooks-like result for graph powers\tnoteref{t1}}
  \tnotetext[t1]{This work was supported by the ANR project HOSIGRA
    (ANR-17-CE40-0022) and by the MUNI Award in Science and Humanities
    of the Grant Agency of Masaryk University.}

\address[1]{Faculty of Informatics, Masaryk University, Botanick\'a 68A, 602 00 Brno, Czech Republic.} 
\address[2]{Former affiliation: Univ. Bordeaux, Bordeaux INP, CNRS, LaBRI, UMR5800, F-33400 Talence, France}
\author[1,2]{Théo Pierron}
\ead{pierron@fi.muni.cz}

\begin{abstract}
  Coloring a graph $G$ consists in finding an assignment of colors
  $c: V(G)\to\{1,\ldots,p\}$ such that any pair of adjacent vertices
  receives different colors. The minimum integer $p$ such that a
  coloring exists is called the chromatic number of $G$, denoted by
  $\chi(G)$. We investigate the chromatic number of powers of graphs,
  i.e. the graphs obtained from a graph $G$ by adding an edge between
  every pair of vertices at distance at most $k$. For $k=1$, Brooks'
  theorem states that every connected graph of maximum degree
  $\Delta\geqslant 3$ excepted the clique on $\Delta+1$ vertices can
  be colored using $\Delta$ colors (i.e. one color less than the naive
  upper bound). For $k\geqslant 2$, a similar result holds: excepted
  for Moore graphs, the naive upper bound can be lowered by 2. We
  prove that for $k\geqslant 3$ and for every $\Delta$, we can
  actually spare $k-2$ colors, excepted for a finite number of
  graphs. We then improve this value to
  $\Theta((\Delta-1)^{\frac{k}{12}})$.
\end{abstract}
\end{frontmatter}

\section{Introduction}

Coloring a graph $G$ consists in finding an assignment of colors
$c: V(G)\to\{1,\ldots,p\}$ such that any pair of adjacent vertices
receive different colors. The minimum integer $p$ such that a coloring
exists is called the chromatic number of $G$, denoted by $\chi(G)$. We
denote by $\Delta(G)$ the maximum degree of the graph $G$. Using a
greedy algorithm, it is easy to show that every graph $G$ can be
colored with $\Delta(G)+1$ colors. A seminal result from Brooks
characterizes the cases when this bound is tight:

\begin{thm}[\cite{brooks1941colouring}]
  Every connected graph $G$ satisfies $\chi(G) \leqslant \Delta(G)$
  excepted when $G=K_{\Delta(G)+1}$ or $\Delta(G)=2$ and $G$ is an odd
  cycle.
\end{thm}

Given an integer $k\geqslant 1$ and a graph $G$, the $k$-th power of
$G$ is the graph obtained from $G$ by adding an edge between vertices
at distance at most $k$ in $G$. In this paper, we are interested in
bounding the chromatic number $\chi(G^k)$ by a function of the integer
$k$ and the maximum degree $\Delta(G)$.

Up to replacing any graph by each of its connected components, we may
assume that all the graphs we consider are connected. Moreover,
coloring powers of paths and cycles is already settled by the
following result.

\begin{prop}[\cite{prowse2003choosability}]
  Let $n$ and $k$ be two integers. Then
  \begin{itemize}
  \item $\chi(P_n^k)=\min(n,k+1)$
  \item If $n>k+1$, then $\chi(C_n^k)=k+1+\lceil\frac{r}{q}\rceil$ where
    $n=q(k+1)+r$ and $r\leqslant k$.
  \item $\chi(C_n^k)=n$ otherwise.
  \end{itemize}
\end{prop}

Therefore, we may only consider graphs with maximum degree at least 3,
so that in particular, none of the graphs we consider in this paper is
a cycle.

For the case of squares of graphs (i.e. $k=2$), we have
$\Delta(G^2)\leqslant \Delta(G)^2$ (and this can be tight). Therefore,
applying Brooks' theorem on $G^2$ states that $\Delta(G)^2$ colors are
sufficient excepted when $G^2$ is a clique on $\Delta(G)^2+1$
vertices. Such graphs are called \emph{Moore graphs}, and Hoffman and
Singleton~\cite{hoffman2003moore} proved that only finitely many of
them exist. For all the other graphs, the $\Delta(G)^2$ bound can
actually be improved, as shown by the following result.

\begin{thm}[\cite{cranston2016painting}]
  \label{thm:sq}
  If $G$ is not a Moore graph, then
  $\chi(G^2)\leqslant \Delta(G)^2-1$.
\end{thm}

\begin{figure}[!ht]
  \centering
  \begin{tikzpicture}[high/.style={inner sep=1.4pt, outer sep=0pt, circle, draw,fill=white},
    level 1/.style={sibling distance = 3cm,level distance = 1.5cm},
    level 2/.style={sibling distance = 1cm,level distance = 1.5cm},
    level 3/.style={sibling distance = 0.33cm,level distance = 1.5cm}] 
    \node[high]  {}
    child{ node [high] {} 
      child{ node [high] {} 
        child{ node [high] {} }
        child{ node [high] {} }
        child{ node [high] {} }
      }
      child{ node [high] {} 
        child{ node [high] {} }
        child{ node [high] {} }
        child{ node [high] {} }
      }
      child{ node [high] {} 
        child{ node [high] {} }
        child{ node [high] {} }
        child{ node [high] {} }
      }
    }
    child{ node [high] {} 
      child{ node [high] {} 
        child{ node [high] {} }
        child{ node [high] {} }
        child{ node [high] {} }
      }
      child{ node [high] {} 
        child{ node [high] {} }
        child{ node [high] {} }
        child{ node [high] {} }
      }
      child{ node [high] {} 
        child{ node [high] {} }
        child{ node [high] {} }
        child{ node [high] {} }
      }
    }
    child{ node [high] {} 
      child{ node [high] {} 
        child{ node [high] {} }
        child{ node [high] {} }
        child{ node [high] {} }
      }
      child{ node [high] {} 
        child{ node [high] {} }
        child{ node [high] {} }
        child{ node [high] {} }
      }
      child{ node [high] {} 
        child{ node [high] {} }
        child{ node [high] {} }
        child{ node [high] {} }
      }
    }
    child{ node [high] {} 
      child{ node [high] {} 
        child{ node [high] {} }
        child{ node [high] {} }
        child{ node [high] {} }
      }
      child{ node [high] {} 
        child{ node [high] {} }
        child{ node [high] {} }
        child{ node [high] {} }
      }
      child{ node [high] {} 
        child{ node [high] {} }
        child{ node [high] {} }
        child{ node [high] {} }
      }
    }

; 
  \end{tikzpicture}
\caption{A $4$-ary tree of height $3$, with $f(4,3)=52$ non-root nodes.}
\label{fig:tree}
\end{figure}
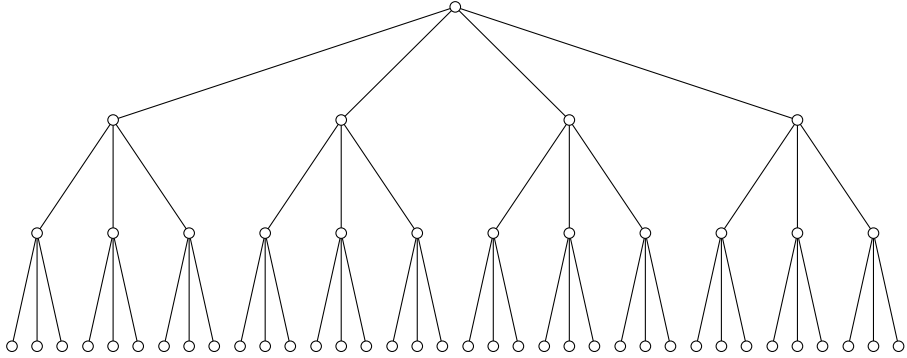

These results have been generalized for higher powers of
graphs. Assume that $k\geqslant 3$. Then, the maximum possible value
of $\Delta(G^k)$ is $f(k,\Delta(G))$, where 
\[f(k,\Delta)=\Delta\sum_{i=0}^{k-1}
  (\Delta-1)^i=\Delta\frac{(\Delta-1)^k-1}{\Delta-2}\] is the number
of nodes of a $\Delta$-ary tree of height $k$, without its root (see
Figure~\ref{fig:tree}).

Brooks' theorem gives that $f(k,\Delta)$ colors are sufficient to
color every graph $G$ with maximum degree $\Delta$, as soon as it is
not a generalized Moore graph, i.e. $G^k$ is not a clique on
$f(k,\Delta)+1$ vertices. However, no such graph exists when
$k\geqslant 3$ (see~\cite{hoffman2003moore}). Therefore, the bound
$\chi(G^k)\leqslant f(k,\Delta)$ always holds. Moreover, this bound is
improved by the following result.
\begin{thm}[\cite{bonamy2014brooks}]
  For $k\geqslant 3$ and every graph $G$, we have
  $\chi(G^k)\leqslant f(k,\Delta)-1$.
\end{thm}

When $k=2$, note that $f(2,\Delta)=\Delta^2$. Hence, together with
Theorem~\ref{thm:sq}, this result settles a conjecture
of~\cite{miao2014distance}, stating that two colors can be spared from
the naive upper bound $f(k,\Delta)+1$, excepted when $k=2$ and $G$ is
a Moore graph.

Bonamy and Bousquet conjecture in~\cite{bonamy2014brooks} that we can
improve this result, by sparing $k$ colors for higher values of $k$,
excepted for a finite number of graphs.

\begin{conj}[\cite{bonamy2014brooks}]
  \label{conj:marthe}
  For every $k\geqslant 2$, only finitely many graphs $G$ satisfy
  $\chi(G^k)\geqslant f(k,\Delta)+1-k$.
\end{conj}

To study how far $\chi(G^k)$ can be from $f(k,\Delta)+1$, we introduce
the following parameter.

\begin{definition}
  For every integer $k$, the \emph{$k$-gap} of a graph $G$, denoted by
  $g_k(G)$, is the difference $f(k,\Delta(G))+1-\chi(G^k)$.
\end{definition}

The aforementioned results can then be rephrased in terms of gap: for
every integer $k\geqslant 2$ and every graph $G$, we have $g_k(G)< 2$
if and only if $k=2$ and $G$ is a Moore graph. Moreover,
Conjecture~\ref{conj:marthe} actually expresses finiteness of the
class of graphs satisfying $g_k<k$. In Section~\ref{sec:k-2}, we prove
the following generic result towards this conjecture.

\begin{thm}
  \label{thm:main}
  Let $k$ and $\Delta$ be two integers with $\Delta\geqslant 3$. Then,
  there are only finitely many graphs~$G$ of maximum degree $\Delta$
  such that $g_k(G)<k-2$.
\end{thm}

Observe that Brooks' theorem gives an infinite list of graphs such
that $g_1<1$, but for every $\Delta\geqslant 3$, this list contains
only one graph of given maximum degree $\Delta$. From that
perspective, Theorem~\ref{thm:main} can be seen as a generalization of
Brooks' theorem for powers of graphs. However, note that the bound
we obtain for $k\leqslant 2$ is worse than the one given by Brooks'
theorem.

In Section~\ref{sec:improve}, we present the following extension of
Theorem~\ref{thm:main}.

\begin{thm}
  \label{thm:improve}
  There exists a function
  $h(k,\Delta)=\Theta((\Delta-1)^{\frac{k}{12}})$ when $k\to\infty$
  such that for all integers $\Delta$ and $k$, only finitely many
  graphs of maximum degree $\Delta$ satisfy $g_k(G)<h(k,\Delta)$.
\end{thm}

Observe that when $k$ and $\Delta$ are large enough, this improved
version implies that there are finitely many graphs with maximum
degree $\Delta$ and satisfying $g_k<k$.  However, this does not imply
Conjecture~\ref{conj:marthe}, since the finiteness result is only
valid when $\Delta$ is fixed.

\section{Proof of Theorem~\ref{thm:main}}
\label{sec:k-2}
In this section, we give a proof of Theorem~\ref{thm:main}. First note
that the cases $k=1,2$ are consequences of Brooks' theorem and
Theorem~\ref{thm:sq}. Thus, we only consider the case $k\geqslant
3$. In the following, we fix two integers $k$ and $\Delta$ greater or
equal to 3, and denote by $G$ a graph of maximum degree $\Delta$. The
proof of Theorem~\ref{thm:main} relies on the following proposition. 
\begin{prop}
  \label{prop:diam}
  If $g_k(G)<k-2$, then the diameter of $G$ is at most $2k-3$.
\end{prop}

Indeed, this ensures that if $g_k(G)<k-2$, then $G$ has at most
$f(2k-3,\Delta)$ vertices. In particular, there are only finitely many
choices for $G$, which concludes the proof of Theorem~\ref{thm:main}.

It now remains to prove Proposition~\ref{prop:diam}. The structure of
the proof is as follows: we assume by contrapositive that $G$ contains
two vertices $a$ and $b$ at distance $2k-2$, and we design an ordering
of the vertices such that the associated greedy procedure gives a
valid coloring of $G^k$ with $f(k,\Delta)+3-k$ colors. As we will see,
this ordering roughly consists in considering the vertices by
decreasing distance to the middle of a fixed shortest path between $a$
and $b$. This shows that $g_k(G)\geqslant k-2$, and proves
Proposition~\ref{prop:diam}.

This section is organized as follows. In a first subsection, we
explain how we count the number of available colors for a given vertex
$v$ when it is considered in the greedy procedure. Then, in the second
subsection, we present the ordering of the vertices and show that each
time a vertex is considered, it has at least an available color. This
ensures that we obtain a valid coloring of~$G^k$ when the procedure
terminates.

\subsection{A framework for counting available colors}

To check whether the coloring procedure we will describe later yields
a valid coloring of $G^k$ with $f(k,\Delta)+3-k$ colors, we need to
count the number of available colors for every vertex $v$ at the
moment it is considered by the procedure. Let $N^k_G(v)$ be the closed
neighborhood of $v$ at distance $k$ in $G$, i.e.
$\{u\in V(G) \mid d(u,v)\leqslant k\}$.

Observe that when $N^k_G(v)$ induces a $\Delta$-ary tree of height $k$
in $G$, the number of uncolored vertices in $N^k_G(v)$ minus $k-3$ is
a lower bound on the number of available colors for $v$.  However,
$N^k_G(v)$ does not necessarily have this nice structure. In
particular, there may be cycles or vertices of degree less than
$\Delta$ in $N^k_G(v)$, and the resulting lower bound is not strong
enough to conclude. We now explain how to deal with this issue.

First, let $T$ be a tree of height $k$, where each node at height less
than $k$ has $\Delta-1$ children. We define $\hat{G}$ as the graph
obtained from $G$ by adding an edge between each vertex $v\in V(G)$
and the root of $\Delta-d_G(v)$ copies of $T$. In particular, observe
that for every $v\in V(G)$, the neighborhood $N_{\hat{G}}^k(v)$
contains only vertices of degree $\Delta$.

\begin{remark}
  In many proofs about graph colorings, we usually remove elements of
  $G$, and use some minimality argument to obtain a coloring to
  extend. In this case, we instead add some vertices. This is not
  related to some inductive argument (we do not even color all of
  these new vertices). The goal of this modification is to obtain a
  better lower bound on the number of available colors for $v$.
  Indeed, we can refine the previous bound by considering the number
  of uncolored vertices in $N^k_{\hat{G}}(v)$ minus $k-3$, which is
  always an improvement since $G^k$ is an induced subgraph of
  $\hat{G}^k$.
\end{remark}

Observe that the number of available colors for $v$ also depends on
whether some colors appear several times in $N^k_{\hat{G}}(v)$, or
similarly if this neighborhood contains cycles. To capture these
cases, instead of counting uncolored vertices, we will consider (in
any fixed order) all the $f(k,\Delta)$ non-empty non-backtracking
walks of length at most $k$ in $\hat{G}$ starting from $v$ (meaning
that we allow the same edge to be used twice, but not in a row).  In
the following, each walk is implicitly assumed to be
non-backtracking. The number of available colors for $v$ thus depends
on the number of such walks ending on an uncolored vertex, or ending
with vertices with the same color (possibly the same vertices).

We say that such a walk is \emph{nice} if its endpoint (possibly $v$)
either is in $V(\hat{G})\setminus V(G)$, or is uncolored, or is the
endpoint of an already considered walk. Each time we find a nice walk
starting at $v$, we say that $v$ \emph{saves a color}. The number of
forbidden colors is the number of non-nice walks. Therefore, the
number of available colors for $v$ is the number of nice walks
starting from $v$, minus $k-3$.

\begin{remark}
  When $N_G^k(v)$ is a $\Delta$-ary tree of height $k$, then each
  non-empty walk corresponds to a unique vertex in
  $N_G^k(v)\setminus\{v\}$. In particular, the nice walks are the ones
  ending at uncolored vertices or at vertices whose color appears at
  the endpoint of an already considered walk. Therefore, counting nice
  walks is actually a generalization of counting uncolored vertices.
\end{remark}

\subsection{The coloring procedure}

We may now give the coloring procedure we consider, and show that it
yields a valid coloring of $G^k$. Since $G$ has diameter at least
$2k-2$, there is a shortest path $P=u_2\cdots u_kxv_1\cdots v_{k-1}$
of length $2k-2$ between two vertices $u_2$ and $v_{k-1}$. First, we
set the color of each $u_i$ (resp. $v_i$) as~$i$. Note that this is a
proper partial coloring of $G^k$: if $d(u_i,v_i)=k$, there is a
path from $u_2$ to $v_{k-1}$ of length $2k-3$, contradicting the
hypothesis.

We fix the following ordering of the vertices of $P$:
$u_2>v_{k-1}>\cdots > u_k >v_1 > x$. Let $w$ be a vertex of
$\hat{G}$. We define the \emph{root} $r_w$ of $w$ as the largest
vertex in $P$ on a shortest path from $w$ to $x$.

Let $N=\{v\in V(\hat{G})\mid r_v=x \text{ and } d(x,v)\leqslant
k\}$. We first color the vertices of $V(G)\setminus N$ by decreasing
distance to $x$, then color the vertices of $V(G)\cap N$, also by
decreasing distance to $x$ (so $x$ will be the last vertex to be
colored). 

\begin{figure}[!ht]
  \centering
  \begin{tikzpicture}[high/.style={inner sep=1.4pt, outer sep=0pt, circle, draw,fill=white},decoration={snake}]
    \node[high,label=below:{$x$}] (x) at (0,0) {};
    \node[high,fill=black,label=below:{$u_k$}] (uk) at (-1,0) {};
    \node[high] (uk') at (-1,1) {3};
    \node[high,fill=black,label=below:{$u_{k-1}$}] (uk1) at (-2,0) {};
    \node[high] (uk1') at (-2,1) {3};
    \node[high,fill=black,label=below:{$u_{i+1}$}] (ui1) at (-4.5,0) {};
    \node[high] (ui1') at (-4.5,1) {3};
    \node[high,fill=black,label=below:{$r_v=u_i$}] (ui) at (-6,0) {};
    \node[high,fill=black,label=below:{$u_1$}] (u1) at (-7,0) {};
    \node[high,fill=black,label=below:{$v_1$}] (v1) at (1,0) {};
    \node[high,fill=black,label=below:{$v_2$}] (v2) at (2,0) {};
    \node[high,fill=black,label=below:{$v_k$}] (vk) at (4,0) {};
    \node[high,label=above:{$v$}] (v) at (-7,4) {};
    \node at (0,1.5) {$N$};
    \node at (0,2.5) {Type 4};
    \node[high] (a) at (-6.25,3) {2};
    \node[high] (b) at (-6,2) {2};

    \node[high] (c) at (-6.75,3) {1};
    \node[high] (d) at (-6.5,2) {1};
    \node[high] (e) at (-6.25,1) {1};
    
    \draw (a) -- (d);
    \draw (b) -- (e);
    \draw (uk1) -- (uk) -- (x) -- (v1) -- (v2);
    \draw[dotted] (u1) -- (ui);
    \draw[dotted] (v2) -- (vk);
    \draw[dotted] (ui1) -- (uk1);
    \draw (ui) -- (ui1) -- (ui1');
    \draw (uk) -- (uk');
    \draw (uk1) -- (uk1');
    \draw (x) -- (-1,3) -- (1,3) -- (x);
     \draw (v) -- (c) -- (d) -- (e);
    \draw[dotted] (e) -- (ui);
  \end{tikzpicture}
  \caption{Global picture of the situation when considering $v$ with
    $r_v\neq x$: black vertices are already colored, white ones are
    uncolored. Integers inside vertices represent types of walks
    ending there.}
  \label{fig:diameter}
\end{figure}
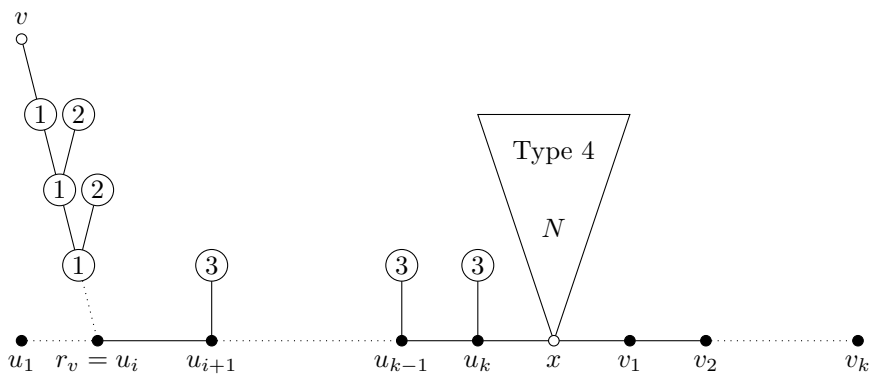

Let $v$ be an uncolored vertex of $G$ outside of $N$. Let
$d=d(v,r_v)$ and $d'=d(r_v,x)$. Denote by
$Q=q_0,q_1,\ldots,q_d$ a shortest path from $v=q_0$ to $r_v=q_d$. We
now consider several types of nice walks in $N^k_{\hat{G}}(v)$, ending on
vertices shown in Figure~\ref{fig:diameter}:
\begin{enumerate}
\item The subpaths of $Q$ ending at some internal vertex $w$ of
  $Q$. By construction, we have
  \[d(w,x)\leqslant d(w,r_v)+d(r_v,x) < d(v,r_v)+d(r_v,x)=d(v,x),\]
  hence $w$ is uncolored when we consider $v$, and the corresponding
  walk is nice. Note that for $j\in [1,d-1]$, there is such a walk of
  length $j$.
\item For each internal vertex $q_j$ of $Q$ with $j>0$, the walk
  $q_0,\ldots,q_j,w'$, where $w'$ is a neighbor of $q_j$ different
  from $q_{j-1}$ and $q_{j+1}$ (which exists since $q_j$ has degree
  $\Delta\geqslant 3$ in $\hat{G}$). Indeed, we have
  \[d(w',x)\leqslant 1+d(q_j,x) \leqslant d(q_{j-1},x) < d(v,x),\]
  hence $w'$ is uncolored when we consider $v$. For every $j\in[3,d]$,
  there is at least one such walk of length $j$.
\item If $r_v=u_i$, the walks $q_0,\ldots,q_{d-1},u_i,\ldots,u_j,u'$,
  where $j\in[i+1,k]$ and $u'$ is a neighbor of $u_j$ in $\hat{G}$
  different from $u_{j+1}$ and $u_{j-1}$. Indeed, we have
  \[d(u',x)\leqslant 1+d(u_j,x) \leqslant d(u_i,x)<d(v,x),\] hence
  $u'$ is uncolored when we consider $v$. For $j\in[d+2,d+d']$, there
  is such a walk of length $j$.
\item If $r_v=u_i$ (hence $v\notin N$), the walks obtained by
  concatenating $Q$ with the subpath of $P$ between $u_i$ and $x$, and
  ending at vertices in $N$ (which are uncolored by
  construction). There is at least one such walk of length $j$ for
  every $j\in [d+d',d+d'+k]$.
\item If $r_v=x$ and $j\in [2,k-1]$, the two walks
  $q_0,\ldots,q_{d-1},x,u_k,\ldots,u_j$ and
  $q_0,\ldots,q_{d-1},x,v_1,\ldots,v_j$ end with vertices sharing the
  same color, hence one of them is nice.
\item If $r_v=x$ and $v\neq x$, then $Q$ is a nice walk since $x$ is
  the last vertex to be colored.
\end{enumerate}

We now have to make sure that at least $k-2$ such walks have length at
most $k$. First consider the case where $v\notin N$. Among the nice
walks of length at most $k$ starting at $v$, the previous counting
ensures that there are:
\begin{itemize}
\item $\min(d-1,k)$ walks of type 1.
\item $\min(d,k)-2$ walks of type 2 if $d\geqslant 2$, and 0
  otherwise.
\item $\min(d+d',k)-d-1$ walks of type 3.
\item $\max(k-d-d'+1,0)$ walks of type 4.
\end{itemize}

If $d\geqslant k$, there are at least $k-1+k-2=2k-3\geqslant k-2$
walks of type 1 and 2, hence we may assume that $d<k$. We separate
four cases depending on whether $d=1$, and whether $d+d'\geqslant k$.

\begin{itemize}
\item If $d=1$ and $d+d'\geqslant k$, then there are at least $k-2$
  walks of type 3.
\item If $d=1$ and $d+d'< k$, then there are at least
  $(d'-1)+(k-d')\geqslant k-2$ walks of type 3 and 4.
\item If $d>1$ and $d+d'\geqslant k$, then there are at least
  \[(d-1)+(d-2)+(k-d-1)=k+d-4\geqslant k-2\]
  walks of type 1, 2, or 3.
\item If $d>1$ and $d+d'< k$, then there are at least
  \[(d-1)+(d-2)+(d'-1)+(k-d-d'+1)=k+d-3\geqslant k-2\]
  walks of type 1, 2, 3 or 4.
\end{itemize}

It now remains to consider the case when $v\in N$. We consider several
cases depending on the value of $d$.
\begin{itemize}
\item If $d\geqslant k$, there are $(k-1)+(k-2)=2k-3\geqslant k-2$ nice
  walks of type 1 and 2.
\item If $\frac{k}{2}\leqslant d<k$, there are
  $(d-1)+(d-2)+1=2d-2\geqslant k-2$ walks of type 1, 2 and 6.
\item If $d <\frac{k}{2}$, observe that there are at least $k-2d$ nice
  walks of type 5 and length at most $k$ (ending either at $u_j$ or
  $v_j$ for each $j\in [d+1,k-d]$). Therefore, there are
  \[(d-1)+(d-2)+(k-2d)+1=k-2\] nice walks of type 1, 2, 5 and 6.
\end{itemize}

Therefore, when considering a vertex $v$, we can always find $k-2$
nice walks of length at most $k$ starting from $v$. This implies that
there is always an available color for $v$ in the greedy procedure,
and $\chi(G^k)\leqslant f(k,\Delta)+3-k$. This concludes the proof of
Proposition~\ref{prop:diam} and thus of Theorem~\ref{thm:main}.

\section{Improving the gap bound}
\label{sec:improve}

This section is devoted to the proof of Theorem~\ref{thm:improve}. We
thus take two integers $k$ and $\Delta$ greater or equal to three, and
$G$ a graph of maximum degree $\Delta$. We again show that if $G$ has
a small $k$-gap, then it should have a small diameter, which ensures
that $G$ can take only finitely many values. More precisely,
Theorem~\ref{thm:improve} relies on the following statement.

\begin{prop}
  \label{prop:improve'}
  For every integer $s\leqslant \frac{k-5}{12}$, if
  $g_k(G)< f(s,\Delta)+1$, then $G$ has diameter at most $k+2s$.
\end{prop}

To obtain Theorem~\ref{thm:improve}, we apply the proposition with
$s=\lfloor\frac{k-5}{12}\rfloor$, and observe that
$f(s,\Delta)+1=\Theta((\Delta-1)^s)=\Theta((\Delta-1)^{\frac{k}{12}})$.

\begin{remark}
  When $k\geqslant 17$, we can take $s=1$ in
  Proposition~\ref{prop:improve'}, an obtain that graphs of maximum
  degree $\Delta$ satisfying $g_k(G)<\Delta+1$ have diameter at most
  $k+2$. In particular, when $\Delta\geqslant k-3$, this strengthens
  Proposition~\ref{prop:diam} by giving more insight on the structure
  of graphs satisfying $g_k(G)<k-2$. Moreover, when
  $\Delta\geqslant k-1$, this also implies a weak version of
  Conjecture~\ref{conj:marthe}: for every
  $\Delta\geqslant k \geqslant 17$, there are only finitely many
  graphs of maximum degree $\Delta$ such that $g_k<k$. This is
  actually valid when $k\geqslant 12$, using some slight optimizations
  in the upcoming proof.
\end{remark}

The end of this section is devoted to the proof of
Proposition~\ref{prop:improve'}. We use similar ideas as for
Proposition~\ref{prop:diam}. Indeed, we assume again that $G$ contains
two vertices $u_1$ and $v_1$ at distance $k+2s+1$ for some
$s\leqslant\frac{k-5}{12}$, and describe a greedy-like procedure
allowing to color $G^k$ with $f(k,\Delta)-f(s,\Delta)$ colors. The
ordering we choose is very similar: we again color vertices by
decreasing distance to the middle of a shortest path, excepted for the
vertices whose root is in the middle, that we color last. The
difference comes from the fact that we do not assign colors to the
vertices of the path beforehand, but only to the endpoints of the path
and their neighbors in $G^s$.

We also reuse part of the framework introduced in the previous
section. More precisely, we construct a graph $\hat{G}$ containing $G$
such that the neighborhood $N_{\hat{G}}^k(v)$ of every vertex
$v\in V(G)$ contains only vertices of degree $\Delta$. We also count
the number of nice walks in order to show that the greedy procedure
does not use more than $f(k,\Delta)-f(s,\Delta)$ colors.

Let $t=\lfloor \frac{k+2s+1}{2}\rfloor$ and
$u_2,\ldots,u_t,v_t,\ldots,v_2$ be the internal vertices of a shortest
path $P$ between $u_1$ and $v_1$ (with $u_t=v_t$ if $k$ is odd).

We first precolor all vertices in
$N_{\hat{G}}^s(u_1)\cup N_{\hat{G}}^s(v_1)$ with colors in
$[1,f(s,\Delta)+1]$. To this end, let
$\gamma_1,\ldots,\gamma_{f(s,\Delta)+1}$ be the walks of length at
most $s$ starting at $u_1$, and color each vertex $w$ in
$N_{\hat{G}}^s(u_1)$ with the smallest index $i$ such that $\gamma_i$
ends in $w$. We apply the same procedure to color
$N_{\hat{G}}^s(v_1)$. The resulting coloring is a proper partial
coloring of $\hat{G}^k$ since otherwise there would be a path of
length at most $k+2s$ between $u_1$ and $v_1$, which is impossible by
hypothesis.

First note that every vertex $v$ satisfying
$N_{\hat{G}}^s(u_1)\cup N_{\hat{G}}^s(v_1)\subset N_{\hat{G}}^k(v)$
saves $f(s,\Delta)+1$ colors. Indeed, for each color
$c\in [1,f(s,\Delta)+1]$, we have two cases:
\begin{itemize}
\item Either the color $c$ appears in both $N_{\hat{G}}^s(u_1)$ and
  $N_{\hat{G}}^s(v_1)$, hence there are two walks starting at $v$ and
  ending on vertices colored with $c$ (one going through $u_1$ and the
  other through $v_1$). This ensures that at least one of these walks
  is nice.
\item Or, by symmetry, the color $c$ does not appear in
  $N_{\hat{G}}^s(u_1)$. Then the endpoint of $\gamma_c$ is also the
  endpoint of some $\gamma_d$ for $d<c$. In particular,
  $\gamma\cup\gamma_c$ is a nice walk starting at $v$.
\end{itemize}
Therefore, there are at least $f(s,\Delta)+1$ nice walks starting at
$v$, i.e. $v$ saves $f(s,\Delta)+1$ colors.

Observe that every vertex in $N_G^s(u_t)\cup N_G^s(v_t)$ is at
distance at most $2s+t\leqslant k$ from every vertex in
$N_{\hat{G}}^s(u_1)\cup N_{\hat{G}}^s(v_1)$. In particular, the
vertices in $N_G^s(u_t)\cup N_G^s(v_t)$ always save $f(s,\Delta)+1$
colors even if they are the last ones to be colored. We may thus focus
on the vertices of $G$ outside of $N_G^s(u_t)\cup N_G^s(v_t)$. We
color them greedily by decreasing distance to $u_t$.

Let $v$ be a vertex of $G$ not in $N_G^s(u_t)\cup N_G^s(v_t)$. We may
assume that $r_v=u_i$ for some $i\in [1,t]$ since the case $r_v=v_i$
is similar. We consider two cases depending on whether the distance
$d=d(v,r_v)$ is small or large.

\begin{itemize}
\item Assume that $d > 3s+t+1$. Let $Q=q_0,\ldots,q_d$ be a
  shortest path between $q_0=v$ and $q_d=u_i$, and let
  $w\in N_{\hat{G}}^s(q_{s+1})$. Then
  \[d(w,u_t)\leqslant d(w,q_{s+1})+d(q_{s+1},u_i)+d(u_i,u_t)\leqslant
    s+d-s-1+d(u_i,u_t)<d+d(u_i,u_t)=d(v,u_t),\] hence $w$ is uncolored
  when we consider $v$, unless $w\in N_G^s(u_1)\cup N_G^s(v_1)$.

  In that case, there is a path of length at most $3s+t+1$ between $v$
  and $u_t$ (going through $q_{s+1}, w$ and $u_1$ or $v_1$). This is
  impossible since
  \[d(v,u_t)=d+t-i> 3s+2t+1-i \geqslant 3s+t+1.\] Therefore, all
  vertices in $N_{\hat{G}}^s(q_{s+1})$ are uncolored when we consider
  $v$. In particular, there are $f(s,\Delta)+1$ nice walks of length
  at most $2s+1\leqslant k$ starting at $v$.

\item Assume that $d\leqslant 3s+t+1$. Let $\ell= \max(2s+2,i+s+1)$
  and $w\in N_{\hat{G}}^s(u_{\min(\ell,t)})$. If $\ell \geqslant t$,
  the vertex $w$ is uncolored by construction. We may thus assume that
  $\ell<t$. In that case, since $i<\ell-s$, we have
  \[d(w,u_t)\leqslant s + d(u_\ell,u_t)= d(u_{\ell-s},u_t)< d(u_i,u_t)=d(v,u_t),\]
  hence again $w$ is uncolored when we consider $v$, unless
  $w\in N_G^s(u_1)\cup N_G^s(v_1)$.

  In that case, observe that there is a path of length at most $2s$
  from $u_1$ (or $v_1$) to $u_\ell$. However, by hypothesis, we have
  $d(u_1,u_\ell)=\ell-1>2s$ and $d(v_1,u_\ell)\geqslant t>2s$, a
  contradiction.

  Therefore, all vertices in $N_{\hat{G}}^s(u_{\min(\ell,t)})$ are
  uncolored when we consider $v$. In particular, there are
  $f(s,\Delta)+1$ nice walks of length at most
  \[d+\min(t,\ell)-i\leqslant d+\ell-i\leqslant \max(5s+t+3-i,4s+t+2)\leqslant 5s+t+2\leqslant
    k\] starting at $v$.
\end{itemize}

In both cases, we can find $f(s,\Delta)+1$ nice walks of length at
most $k$ starting at $v$, i.e. $v$ saves $f(s,\Delta)+1$ colors. This
implies that $g_k(G)\geqslant f(s,\Delta)+1$, which ends the proofs of
Proposition~\ref{prop:improve'} and Theorem~\ref{thm:improve}.

\section{Conclusion}

Theorem~\ref{thm:main} is a first step towards
Conjecture~\ref{conj:marthe}. A very similar proof allows us to
replace $k-2$ by $k-1$ when $\Delta\geqslant 4$. However, even for
large $\Delta$, the current proof cannot yield any better bound on
$g_k$ (say, $k$). Indeed, the bottleneck is reached by vertices in
$N_G(x)\setminus V(P)$ whose neighbors in $G^k$ (excepted $x$) are all
colored and only the colors $[2,k-1]$ appear twice in their
neighborhood at distance $k$. Therefore, these vertices can save at
most $k-1$ colors in the worst case regardless of the value of
$\Delta$.

Observe also that while bounding the diameter of graphs with $g_k<k-2$
is sufficient to obtain Theorem~\ref{thm:main}, we would need more
insight on their structure to prove Conjecture~\ref{conj:marthe}. It
is not hard to extend the previous methods to show that such graphs
are $\Delta$-regular, and that their girth is at least $k+2$.

Regarding Theorem~\ref{thm:improve}, there is no reason to believe
that $\Theta((\Delta-1)^{\frac{k}{12}})$ is the largest possible value
of $g$ such that for every integers $k$ and $\Delta$, there are only
finitely many graphs of maximum degree $\Delta$ such that $g_k<g$.
Indeed, it may be possible to improve this value by considering a more
involved coloring procedure. For example, trying to merge the
procedures we presented in Propositions~\ref{prop:diam}
and~\ref{prop:improve'} may lead to a better bound. However, no
generic upper bound seem to be known. This leads to the following
question.

\begin{question}
  For all integers $k$ and $\Delta$, what is the smallest $g$ such
  that infinitely many graphs of maximum degree $\Delta$ satisfy
  $g_k(G)=g$?
\end{question}

\bibliographystyle{plain}

\end{document}